\documentclass[a4paper,twoside,10pt]{article}

\input{jgrg19.sty}
\begin{document}
%
\pagestyle{fancy}
\fancyhead{}
  \fancyhead[RO,LE]{\thepage}
  \fancyhead[LO]{C. Bambi}                  
  \fancyhead[RE]{Accretion process in Kerr space-times}    
\rfoot{}
\cfoot{}
\lfoot{}
\label{O36}    
\title{%
  Numerical simulations of the accretion process in Kerr space-times 
  with arbitrary value of the Kerr parameter
}
%
\author{%
  Cosimo Bambi\footnote{Email address: cosimo.bambi@ipmu.jp}
}
%
\address{%
  Institute for the Physics and Mathematics of the Universe,\\
  The University of Tokyo,\\ 
  Kashiwa, Chiba 277-8568, Japan}
%
\abstract{
According to the Cosmic Censorship Conjecture, all the singularities
produced by the collapsing matter must be hidden behind an event 
horizon. In 4D general relativity, this implies that the final product 
of the collapse is a Kerr-Newman black hole. Here I consider the 
possibility that the Cosmic Censorship Conjecture can be violated. 
I present the results of some numerical simulations of the accretion 
process onto Kerr black holes (objects with event horizon) and Kerr 
super-spinars (fast-rotating objects without event horizon). This is 
a preliminary study to investigate how the Cosmic Censorship Conjecture 
can be tested by astrophysical observations.   
}

\section{Introduction}
Today gravity is relatively well tested in the weak field limit,
while little or nothing is known when it becomes strong~\citep{O36_will}. 
Strong gravitational fields can be found around astrophysical compact 
objects and could be probed by studying the radiation emitted in 
the accretion process. However, that turns out to be a very difficult 
job: the radiation emitted by the falling gas depends significantly 
on the accretion model and it is apparently impossible to constrain 
gravity without several model-dependent assumptions. For such a
reason, today we know some ``black hole candidates'', but actually 
we do not know if these objects have an event horizon or if the 
space-time around them is described by the Kerr metric. In astrophysics, 
one assumes that these candidates are Kerr black holes and studies 
different scenarios of accretion in order to explain observations. 
Here I am instead interested in testing the actual nature of these
objects. In particular, I consider the possibility that some black
hole candidates rotate too fast to have an event horizon and I show 
that the accretion process onto them would be so much different 
that hopefully future theoretical studies and astrophysical observations 
will be able to confirm or rule out such a possibility.

\section{The Cosmic Censorship Conjecture}
In general relativity, under apparently reasonable assumptions, the 
collapsing matter leads inevitably to the formation of singularities. 
Here there are two possibilities: ${\it i)}$ the singularity is hidden 
behind an event horizon and the final product is a black hole, 
${\it ii)}$ the singularity is not hidden behind an event horizon and 
therefore is naked. Since space-times with naked singularities typically
have pathologies, usually some form of the Cosmic Censorship Conjecture 
is assumed and naked singularities are forbidden~\citep{O36_penrose}. 
Neglecting the electric charge, it turns out that, in four dimensions, 
the final product of the gravitational collapse of matter is a Kerr black 
hole~\citep{O36_carter, O36_robinson}.

The Kerr metric is completely characterized by two parameters; that is, 
the mass $M$ and the spin $J$. The latter is often replaced by the Kerr 
parameter $a$, defined as $a = J/M$. Using Boyer-Lindquist coordinates, 
the position of the horizon of a Kerr black hole is given by
\begin{eqnarray}
r_H = M + \sqrt{M^2 - a^2} \, ,
\end{eqnarray}
which demands the well known constraint $|a| \le M$. For $|a| > M$,
there is no horizon and the space-time contains a naked singularity. In absence
of horizon, it is possible to reach the physical singularity at $r = 0$ from 
some large $r$ in finite time, enter the ring singularity, go to the region 
with negative values of $r$, and eventually come back to the starting point 
at an earlier time. So, the theory allows for the existence of closed 
time-like curves and causality can be violated.

However, it is widely believed that the Planck scale, 
$E_{Pl} \sim 10^{19}$~GeV, is the natural UV cut-off of classical general 
relativity. In other words, the theory would be unable to describe 
phenomena with a characteristic energy exceeding $E_{Pl}$. If we apply 
this general idea to the case of the Kerr space-time with $|a| > M$, where 
observer-independent quantities like the scalar curvature diverge at the 
singularity, it is at least questionable to expect that the prediction of 
the existence of closed time-like curves is reliable. New physics could 
instead replace the singularity with something else and Nature may conserve 
causality, not because it is impossible to create an object with $|a| > M$,
but because there is no singularity in the full theory. On the basis of 
this argument, super-spinning Kerr objects with no event horizon, or 
``super-spinars'', might exist in the Universe~\citep{O36_gimon}.

\section{Accretion process}
\subsection{Model and assumptions}

The first step to study the radiation emitted in the accretion process 
onto a compact object is to investigate the accretion process itself.
In Ref.~\citep{O36_sim}, I discussed the accretion process of a test
fluid in a background Kerr space-time; that is, I neglected the
back-reaction of the fluid to the geometry of the space-time, as well
as the increase in mass and the variation in spin of the central object 
due to accretion. Such an approximation is surely reasonable to describe 
the accretion onto a stellar mass compact object in a binary system, because 
in this case the matter captured from the stellar companion is typically 
small in comparison with the total mass of the compact object. The results 
of these simulations should instead not be applied to long-term accretion 
onto a super-massive object at the center of a galaxy, where accretion 
makes the mass of the compact object increase by a few orders of magnitude 
from its original value.

The calculations are made with the relativistic hydrodynamics module
of the public available code PLUTO~\citep{O36_mignone}, properly modified 
for the case of curved space-time, as described in~\citep{O36_sim}. The 
computational domain is the 2D axisymmetric space $r_{in} < r < 20 \, M$
and $0 < \theta < \pi$, where $r_{in}$ is set just outside the event
horizon in the case of black hole, and $r_{in} = 0.5 \, M$ in the case of 
super-spinar. The choice of $r_{in} = 0.5 \, M$ may appear arbitrary, but 
it was checked that does not significantly alter the final result for any 
value of $|a|/M$, as long as $r_{in} \lesssim 0.7 \, M$.

Here the accretion process is spherically symmetric and the gas is injected
from the outer boundary at a constant rate\footnote{Spherical or quasi-spherical
accretion flows are expected when the compact object accretes from the
stellar medium or when it belongs to a binary system in which the companion
is massive and has a strong stellar wind.}. Because of the simple treatment
of the accreting matter, the gas temperature is not under control.
In~\citep{O36_sim} I simply imposed a maximum temperature: the aim was not
to find an accurate description of the accretion process, but to catch
some peculiar features of the accretion process onto Kerr objects with
$|a| > M$. The code was run with $T_{max} =$ 10~keV, 100~keV, and 1~MeV,
obtaining essentially the same result. Such a range of $T_{max}$ is
the one suggested by observations of galactic black hole candidates: the
hard X-ray continuum (10 -- 200~keV) is a typical feature of all these 
objects and is often explained with a hot inner disk or a hot corona, in
which the electron temperature is around 100~keV (see e.g. 
Ref.~\citep{O36_liang}). Let us notice that in this case the accretion 
process is not the simple Bondi accretion. In the Bondi accretion, the 
temperature of the gas (ions) at the horizon (in the case of black holes) 
is about 100~MeV and the proper velocity of the flow is close to 1.

\subsection{Results}
\begin{figure}[t]
\centering
\includegraphics[keepaspectratio=true,width=7cm]{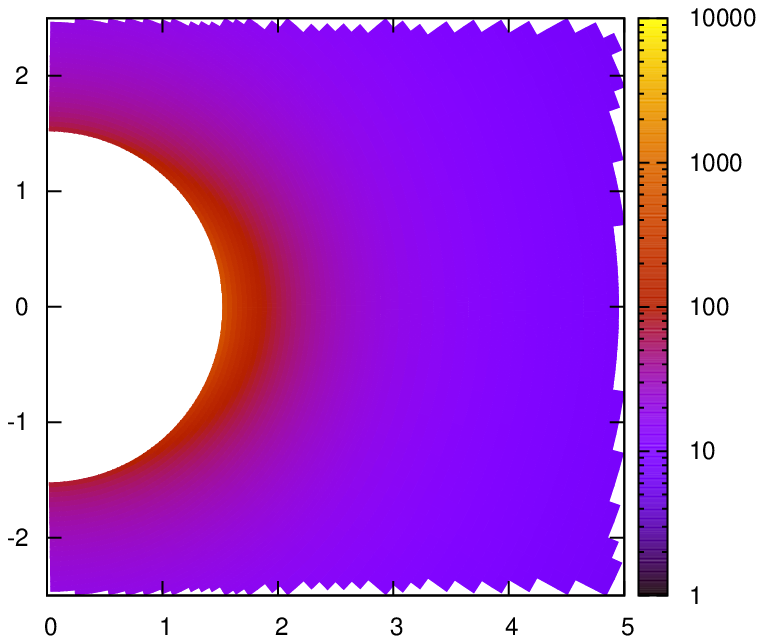}
\includegraphics[keepaspectratio=true,width=7cm]{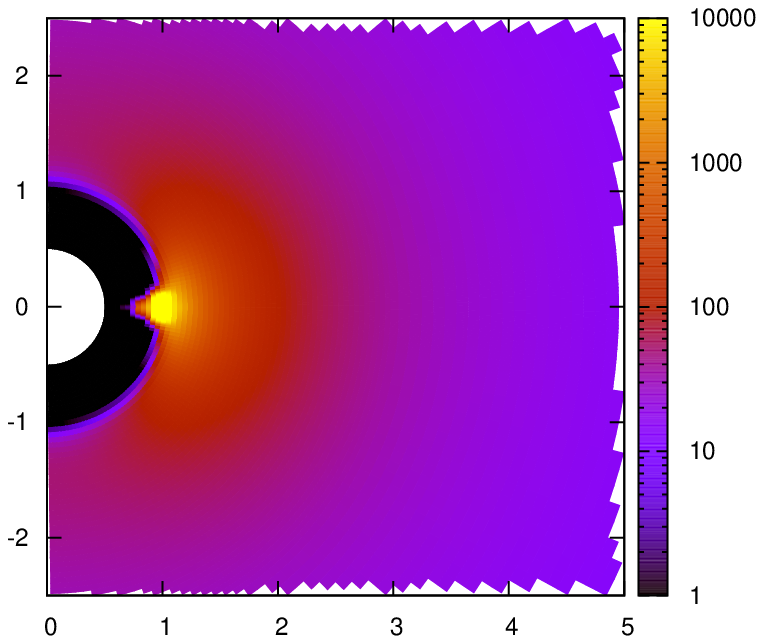} \\
\includegraphics[keepaspectratio=true,width=7cm]{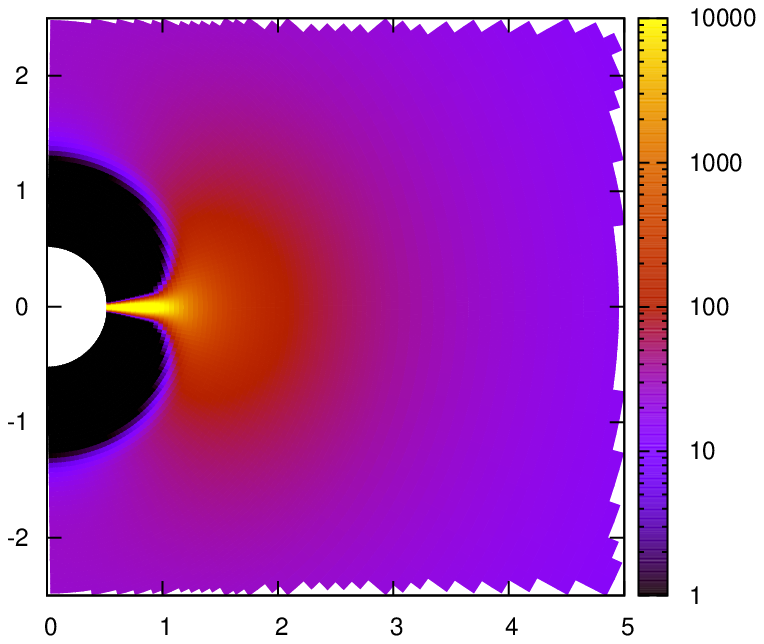}
\includegraphics[keepaspectratio=true,width=7cm]{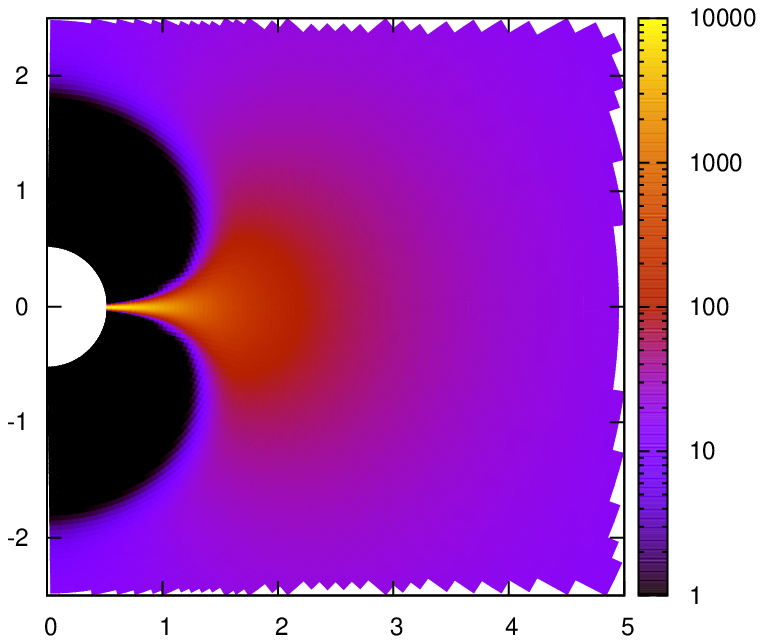}
\caption{
Density plot of the accretion flow around a Kerr BH with $a_* = 0.9$
(top left panel) and super-spinars with $a_* = 1.1$ (top right panel),
$a_* = 1.4$ (bottom left panel), and $a_* = 2.0$ (bottom right panel).
The density scale (shown on the right hand side of each panel) is in
arbitrary units. The unit of length along the $x$ and $y$ axes is $M$.
The white area is out of the domain of computation. The peculiar
feature of the super-spinar case is that most of the space-time around 
the central object is almost empty (the black regions in the pictures):
that is the result of the repulsive force at short distance from the 
center. For more details, see Ref.~\citep{O36_sim}.
}
\label{fig:O36_sim}
\end{figure}

The results of the simulations are summarized in Fig.~\ref{fig:O36_sim},
where it is shown the rest-mass energy density of the accretion flow
around a Kerr black hole with $a/M = 0.9$ (top left panel) and super-spinars 
with $a/M = 1.1$ (top right panel), 1.4 (bottom left panel), and 2.0 
(bottom right panel). The peculiar feature of the case with $|a|/M > 1$ is 
that the gravitational force near the massive object can be repulsive. 

There are three qualitatively different cases determined by the value of 
$|a|/M$ (for more details, see Ref.~\citep{O36_sim}): \\
$1)$ Black hole with $|a|/M \le 1$. We have the usual accretion picture:
the injected matter always reaches a quasi-steady state configuration, in
which matter is lost behind the event horizon at the same rate as it 
enters the computational domain. \\
$2)$ Super-spinars with $|a|/M > 1$. The gravitational force in the
neighborhood of the center $r = 0$ can be repulsive (because of the singularity,
not of the rotation of the gas) and thus makes the accretion process harder.
The critical radius where the gravitational force changes from attractive 
to repulsive can be estimated analytically. In Boyer-Lindquist coordinates, 
it is roughly determined by the sign of the quantity $r^2 - a^2 \cos^2\theta$, 
i.e. the force is attractive (repulsive) if $r^2 - a^2 \cos^2\theta > 0$ 
($< 0$), where $\theta$ is the polar angle\footnote{This simple criterion
works better for higher values of $|a|/M$ and cannot explain some important
features of the case $1 < |a|/M < 1.4$.}. There are two different super-spinar
regimes: \\
$2a)$ Super-spinars with $|a|/M < 1.4$. The accretion process is extremely
suppressed and only a small amount of the accreting gas can reach the center.
Most of the gas is accumulated around the object, forming a high density
cloud that continues to grow. Apparently, no quasi-steady state exists. \\
$2b)$ Super-spinars with $|a|/M \ge 1.4$. In this case the repulsive force is
not capable of preventing a regular accretion of the object. The flow reaches
the center by forming a sort of high density disk on the equatorial plane.
A quasi-steady state configuration is possible.

The origin of the critical value $|a|/M = 1.4$ can be understood with an
analytical argument, as the maximum value for which there are stable marginally
bound orbits on the equatorial plane with zero angular momentum at infinity.

\section{Final remarks}
It is widely believed that the final product of the gravitational collapse
of matter is a Kerr black hole. However such a conclusion is based on a set
of unproved assumptions, including the Cosmic Censorship Conjecture. In this
talk I discussed the possibility that the Cosmic Censorship Conjecture can
be violated and I presented some astrophysical implications. I showed that
the accretion process in Kerr space-time onto objects with event horizon
(black holes) and without event horizon (super-spinars) is quite different 
and hopefully the two scenarios can be distinguished observationally with no 
model-dependent assumptions.

Two comments are in order here. First, if the Cosmic Censorship Conjecture
is violated, there is no uniqueness theorem guaranteeing that the space-time
is described by the Kerr metric. And indeed other axisymmetric solutions of 
the Einstein equations in vacuum with naked singularities are known (e.g. the
Weyl space-times). Here I discussed the case of super-spinar because it is
likely the simplest example. Second, we do not know if super-spinars are
stable objects and this question is presumably difficult to address, because 
we do not know how the singularity is solved in the full theory.

The study of the accretion process onto objects with and without event horizon
is the first step to figure out possible observational signatures of the
radiation emitted in the accretion process which can be used to test the Cosmic
Censorship Conjecture. For example, an application of this work will be the
predictions of the ``direct image'' of super-spinars~\citep{O36_shadow1, O36_shadow2}.

\section*{Acknowledgments}
I wish to thank my collaborators, Katherine Freese, Tomohiro Harada, 
Rohta Takahashi, and Naoki Yoshida.
This work was supported by World Premier International Research Center 
Initiative (WPI Initiative), MEXT, Japan.


\end{document}